\documentclass[twocolumn,showpacs,preprintnumbers,amsmath,amssymb]{revtex4}

\usepackage{graphicx}
\usepackage{dcolumn}
\usepackage{bm}

\begin{document}

\title{Universal Scaling Behavior of Clustering Coefficient\\ Induced by Deactivation Mechanism}
\author{Liang Tian}
\author{Chen-Ping Zhu}
 \email{chenpingzhu@yahoo.com.cn}
\author{Da-Ning Shi}
 \email{shi@nuaa.edu.cn}
\author{Zhi-Ming Gu}
\affiliation{College of Science, Nanjing University of Aeronautics
and Astronautics, Nanjing, 210016, PR China}
\author{Tao Zhou}
\affiliation{Nonlinear Science Center and Department of Modern
Physics, University of Science and Technology of China, Hefei,
Anhui, 230026, PR China}

\date{\today}

\begin{abstract}
We propose a model of network growth that generalizes the
deactivation model previously suggested for complex networks.
Several topological features of this generalized model, such as the
degree distribution and clustering coefficient, have been
investigated analytically and by simulations. A scaling behavior of
clustering coefficient $C \sim 1/M$ is theoretically obtained, where
$M$ refers to the number of active nodes in the network. We discuss
the relationship between the recently observed numerical behavior of
clustering coefficient in the coauthor and paper citation networks
and our theoretical result. It shows that both of them are induced
by deactivation mechanism. By introducing a perturbation, the
generated network undergoes a transition from large- to small-world,
meanwhile the scaling behavior of $C$ is conserved. It indicates
that $C \sim 1/M$ is a universal scaling behavior induced by
deactivation mechanism.
\end{abstract}

\pacs{89.75.Hc, 87.23.Ge, 89.65.-s, 89.75.Fb}

\maketitle

\section{INTRODUCTION}
Many social, biological, and communication systems can be properly
described as complex networks with nodes representing individuals or
organizations and links mimicking the interactions among them[1-3].
Examples are numerous: these include the Internet[4,5], the World
Wide Web[6,7], biological networks[8,9], food webs[10], social
networks[11], etc. Recent empirical studies indicate that the
networks in various fields exhibit some common topological
characteristics: a small average distance as random networks, large
clustering coefficient as regular networks (small-world
property)[12] and a power-law degree distribution (scale-free
property)[13]. The ubiquity of complex networks has inspired
tremendous investigations on them. Among these flourishing
researches, the effect of aging is of particular interest[14-18],
since it is a universal mechanism in reality. For instance, in the
movie actor collaboration network, the more famous an actor is, the
more chances he will have to act in new movies. But, no matter how
famous he may be, every star will become gradually inactive as time
goes on. This aging effect can greatly influence the evolution of
networks and results in peculiar network structural property[14,15].

Recently, B\"orner \emph{et al}. introduced a general process model
that simultaneously grows coauthor and paper citation networks[19],
in which the core assumption is that the twin networks of scientific
researchers and academic articles mutually support one another. In
their model, each of the authors and papers is assigned a topic, and
authors read, cite, produce papers or coauthor with others only in
their own topic area. Interestingly, they found that the clustering
coefficient $C$ of the simulated paper citation network is linearly
correlated with the number of topics. We note that the main
underlying dynamic rule governing the evolution of the network is
aging. For example, due to the lifespan of human, once authors are
older than a specified age, they will be set deactivated, and do not
produce papers or coauthor with others any longer. Furthermore,
papers cease to receive links when their contents are outdated.
Therefore, these considerations motivate us to theoretically
investigate the effect of aging on the clustering coefficient of the
network. In the present paper, we concentrate on this ingredient of
self-organization of the coauthor and paper citation networks and
propose a simple generalized model, in which the main dynamic is
deactivation mechanism. We will demonstrate that the behavior of
clustering coefficient $C$ in the coauthor and paper citation
networks is universal in networks generated by deactivation
mechanism.

This paper is organized as follows. In section II, the model is
introduced. In section III, we give both the numerical and analytic
results about the effect of deactivation mechanism on network
structure, including degree distribution (Sec. III A) and clustering
coefficient (Sec. III B). An interesting scaling behavior of $C$ is
obtained. In section IV, a structural perturbation is introduced. We
show that the perturbation leads to a structural transition from
large- to small-world (Sec. IV A), while the scaling behavior of $C$
is conserved (Sec. IV B). Finally in section V, we discuss the
relationship between our result and the behavior of clustering
coefficient in the coauthor and paper citation networks and give a
summary .

\section{THE MODEL}
In the present model, each node can be in two different states:
active or inactive[18,20,21]. The evolution process starts with a
one-dimensional lattice consisting of $M$ active nodes with periodic
boundary condition and coordination number $2z$[22]. Then, in each
time step

\

(1) Add a new node into the network, and connect it to $m$ nodes
randomly chosen from the $M$ active ones.

(2) Activate the new node.

(3) Deactivate one of the active nodes. The probability that the
node $i$ is deactivated is given by
\begin{eqnarray}\pi(k_i)=\frac{\alpha}{k_i},\end{eqnarray}
where the normalization factor is defined as
$\alpha=(\sum_{j\in\mathcal{A}}1/k_j)^{-1}$. The summation runs over
the set $\mathcal{A}$ of the currently $M+1$ active nodes.

\

It is worthwhile to note that, when $M=m$ and $z=[\frac{M}{2}]$, the
present model reduces to the famous deactivation model introduced by
Klemm and Egu\'iluz (KE model)[18]. For convenience, we call this
generalized deactivation model GKE model.

\section{STRUCTURAL PROPERTIES}
\subsection{Degree distribution}
By using the continuous approximation similar to that used in Ref.
[18], the degree distribution $P(k)$ can be obtained analytically
for GKE model. Let us first derive the degree distribution
$p^{(t)}(k)$ of the active nodes at time $t$. It evolves according
to the following master equation:
\begin{eqnarray}p^{(t+1)}(k+1)&=&p^{(t)}(k)\frac{m}{M}[1-\pi(k)] \nonumber \\ &+&p^{(t)}(k+1)[1-\frac{m}{M}][1-\pi(k+1)].\end{eqnarray}
On the right side of Eq. (2), the first term accounts for the
process in which an active node with degree $k$ at time $t$ is
connected to the new node and not deactivated in the next time step;
The second term indicates the process that an active node with
degree $k+1$ at time $t$ is not connected to the new node and still
active in the next time step.

\begin{figure}
\scalebox{0.9}[0.9]{\includegraphics{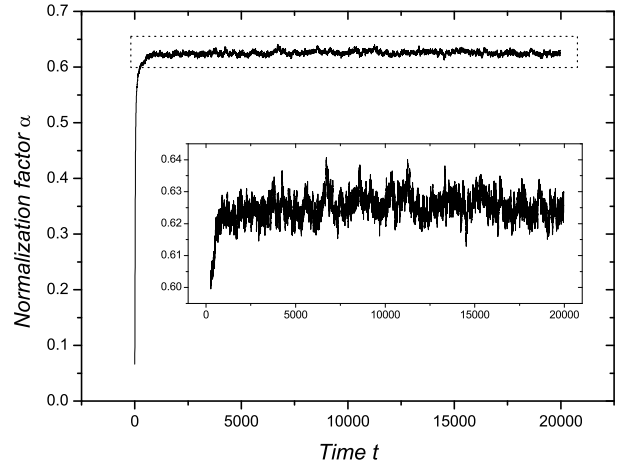}} \caption{Illustration of
the normalization factor $\alpha$ as a function of time $t$ with the
parameters $m=10$ and $M=30$. The amplified version can be seen in
the inset. The data points correspond to system size
$N=2\times10^4$, and each is obtained as an average of $100$
independent runs.}
\end{figure}

We investigate the behavior of $\alpha$ in time evolution. Fig. 1
shows the dependence of the normalization factor $\alpha$ on time
$t$. We find that $\alpha$ approaches a stable value with certain
fluctuations as soon as the evolution of the network starts. We
assume that the fluctuations of the normalization factor $\alpha$
are small enough, i. e., it can be treated as a constant. Then, the
stationary case $p^{(t+1)}(k)=p^{(t)}(k)$ of Eq. (2) yields
\begin{eqnarray}p(k+1)-p(k)=\frac{-\alpha-(\gamma-1)k}{k(k+\gamma-\alpha)}p(k),\end{eqnarray}
where $\gamma=\frac{\alpha M}{m}+1$. Treating $k$ as continuous we
write down the equation
\begin{eqnarray}\frac{dp}{dk}=\frac{-\alpha-(\gamma-1)k}{k(k+\gamma-\alpha)}p(k),\end{eqnarray}
which yields the solution
\begin{eqnarray}p(k)\sim k^{-\gamma+1}.\end{eqnarray}
When the system size $N$ is large compared with $M$, the degree
distribution of the whole network $P(k)$ can be approximated by
considering the inactive nodes only. Thus $P(k)$ can be calculated
as the rate of the change of the degree distribution $p(k)$ of the
active nodes. We find
\begin{eqnarray}P(k)=-\frac{dp}{dk}=ck^{-\gamma},\end{eqnarray} where
$c=(\gamma-1)m^{\gamma-1}$ is the normalization constant. Finally,
the exponent $\gamma=3$ is obtained from a self-consistent condition
\begin{eqnarray}2m=\int_m^\infty k P(k)dk.\end{eqnarray}

The exponent $\gamma$ can be tunable if we introduce the initial
attractiveness just like that of the model in Ref. [18]. Since it is
not our focus, we will not show this effect here.

In Fig. 2, we plot the cumulative degree distribution of GKE
networks by simulations. We obtain a power law scaling with
best-fitted exponent $\gamma-1=1.96\pm0.02$, which is in agreement
with the analytical result. In fact, the exponent $\gamma$ is
dependent on $m$[20], which can be ignored when $m$ is large.
However, the number of active nodes $M$ has no effect on degree
distribution exponent $\gamma$, which is analytically and
numerically obtained.

\begin{figure}
\scalebox{0.8}[0.8]{\includegraphics{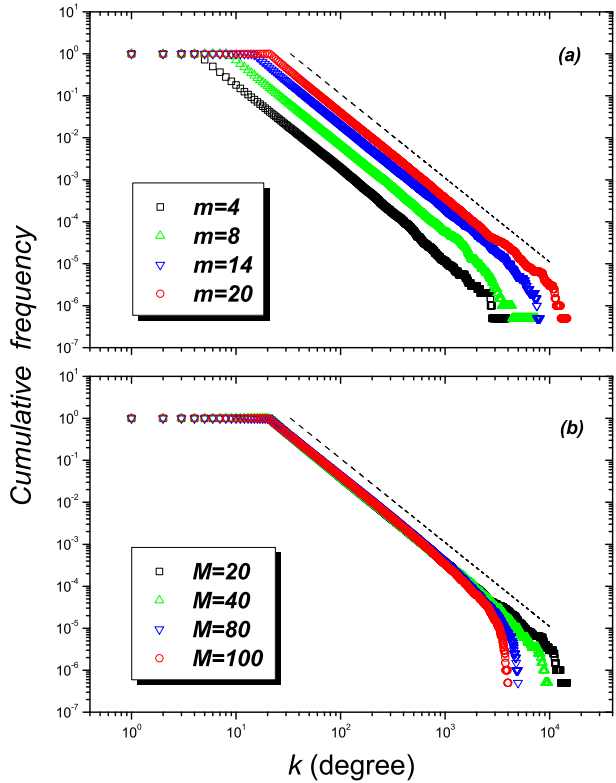}} \caption{(Color online)
Cumulative degree distribution of GKE networks with parameters (a)
$M=20$; $m=4$ (squares), $8$(upward triangles), $14$ (downward
triangles), $20$ (circles) and (b) $m=20$; $M=20$ (squares),
$40$(upward triangles), $80$ (downward triangles), $100$ (circles).
The data points correspond to system size $N=2\times10^4$, and each
is obtained as an average of $100$ independent runs. The two dash
lines have slope $-2.0$ for comparison.}
\end{figure}

\subsection{Clustering coefficient}
The clustering coefficient $C(l)$ of node $l$ with degree $k_l$ can
be defined as follow:
\begin{eqnarray}C(l)=\frac{2E(l)}{k_l(k_l-1)},\end{eqnarray} where $E(l)$ is the number
of links between neighbors of node $l$.

According to the definition of the GKE model, when a new node with
$m$ links is added into the network, the links are attached to the
nodes randomly selected from the active ones. Thus, the probability
that two arbitrary active nodes are connected is $\frac{m}{M}$. It
follows that a node $l$ with degree $k_l=m$ has
\begin{eqnarray}E(l)=\frac{m}{M}\frac{k_l(k_l-1)}{2}.\end{eqnarray}
If $l$ is deactivated in the time step of its generation its
neighborhood does not change any more and $C(l)$ keeps stable.
Otherwise, node $l$ is not deactivated. In the next time step, a new
node $j$ is added. As we note, the probability that node $j$ makes
connection to $l$ is equal to the probability that one of the
neighbors of node $l$ is deactivated in the last time step. We
assume that if $k_l$ is added by $1$, one of its active neighbors
has already been deactivated in the last time step. Thus, when the
newly added node is connected to node $l$, one of its neighbors $s$
is inactive and one possible link between the newly added node and
$s$ is missed. Then we have
\begin{eqnarray}E(l)=\frac{m}{M}[\frac{k_l(k_l-1)}{2}-1],\end{eqnarray}
where $k_l=m+1$. Also, if $k_l=m+2$, there will be $2$ inactive
nodes in the neighbors of node $l$ causing another $2$ possible
links to be missed. Thus we obtain
\begin{eqnarray}E(l)=\frac{m}{M}[\frac{k_l(k_l-1)}{2}-1-2],\end{eqnarray}
where $k_l=m+2$. This process repeats until node $l$ is deactivated,
whose neighborhood does not change any more. By induction, we have
\begin{eqnarray}E(l)=\frac{m}{M}[\frac{k_l(k_l-1)}{2}-\sum_{\nu=1}^{k_l-m}\nu].\end{eqnarray}
Thus the clustering coefficient $C(l)$ depends only on the degree
$k_l$. The exact relation is
\begin{eqnarray}C(l)=\frac{m}{M}[1-\frac{(k_l-m+1)(k_l-m)}{k_l(k_l-1)}].\end{eqnarray}
The clustering coefficient $C$ of the whole network is the average
of $C(l)$ over all nodes, i. e.,
\begin{eqnarray}C=\frac{1}{N}\sum_{l=1}^N \frac{m}{M}[1-\frac{(k_l-m+1)(k_l-m)}{k_l(k_l-1)}].\end{eqnarray}
Writing Eq. (14) in continuous form yields
\begin{eqnarray}C=\int_m^\infty \frac{m}{M}[1-\frac{(k-m+1)(k-m)}{k(k-1)}]P(k)dk,\end{eqnarray}
where $P(k)$ is the degree distribution which we have derived above.
Finally, the result is
\begin{eqnarray}C=\frac{1}{M}[\frac{5m}{6}-\frac{7}{30}+\mathcal{O}(m^{-1})].\end{eqnarray}
Obviously, when $M=m$, the clustering coefficient of the KE model is
recovered[23].

From Eq.(16), we know that the clustering coefficient $C$ is
independent of the system size $N$. This asymptotic behavior of $C$
is reported in Fig. 3. In the limit of large $N$, the clustering
coefficient $C$ gets to an stationary value of $0.31$, which agrees
with the analytical result.

It is important to point out that the clustering coefficient has an
novel scaling behavior $C\sim1/M$. Extensive numerical simulations
perfectly confirm this result (see Fig. 4). This behavior can be
related to the recent numerical study on the coauthor and paper
citation networks, which will be discussed in Sec. V.

\begin{figure}
\scalebox{0.9}[0.9]{\includegraphics{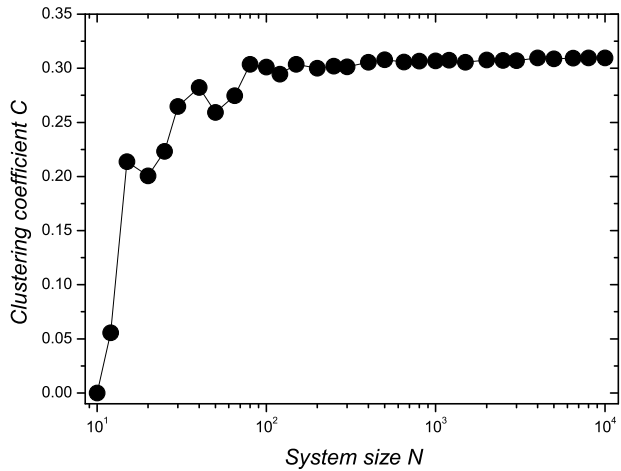}} \caption{Illustration of
the average clustering coefficient $C$ as a function of system size
$N$ with the parameter $m=4$ and $M=10$. The clustering coefficient
$C$ approaches a stationary value about 0.31, which is precisely
predicted by Eq. (16). Each data point is obtained as an average of
$1000$ independent runs.}
\end{figure}

\begin{figure}
\scalebox{0.9}[0.9]{\includegraphics{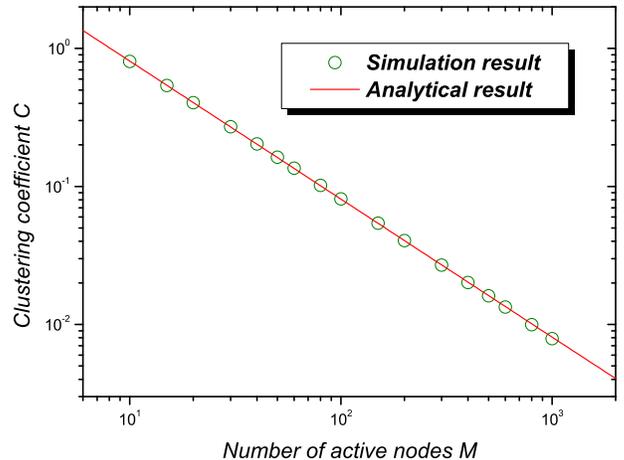}} \caption{(Color online)
The analytical result of the clustering coefficient of GKE network,
$C(M)$, as a function of $M$ (line), in comparison with the
simulation (circles) results. Other parameters for the simulation
are $m=10$ and $N=2\times10^4$. Each data point is obtained as an
average of $100$ independent runs.}
\end{figure}

\section{STRUCTURAL PERTURBATION}
\subsection{Structural transition}
We introduce a structural perturbation to the GKE model by modifying
step (1) of the definition as follow: Add a new node with $m$ links
to the network. With probability $p$, attach {\em one} of the new
node's links to a randomly selected {\em inactive} node. The other
links are then attached to nodes chosen randomly from the $M$ {\em
active} ones. We will show that the perturbation will lead to a
phase transition[24] from large- to small-world in the network
without changing the scale-free property.

In GKE model, each node can be represented by the time step of its
generation. It is clear that, when $p=0$ the GKE network is
structured[18], i. e., the time ordering exists and the mean field
manner is absent[20,25]. We denote $l(t)$ as the average distance
for pairs of nodes separated by time interval $t$. Fig. 5 shows the
simulation results of the variation of $l(t)$ with perturbation
parameter $p$. It can be found that, when $p=0$, $l(t)$ increases
linearly with $t$, i. e., the time ordering indeed exists. Since the
nodes in the network are uniformly distributed on time axis, we can
easily obtain that the average distance $L$ is linearly correlated
to the system size $N$, i. e., $L\propto N$, which indicates the
absence of small world effect. However, once $p$ is a small finite
value, $l(t)$ becomes independent of time interval $t$, i. e., the
time ordering vanishes. Meanwhile, all nodes with the same degree
can be considered to be statistically equivalent, and the mean-field
manner is recovered.

\begin{figure}
\scalebox{0.9}[0.9]{\includegraphics{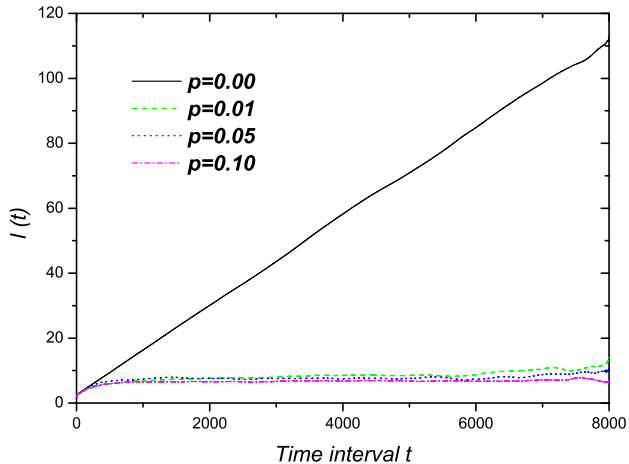}} \caption{(Color online)
Illustration of $l(t)$ as a function of time interval $t$, with
perturbation $p=0.00$ (solid line), $0.01$ (dashed line), $0.05$
(dotted line), and $0.10$ (dot-dashed line). Other parameters for
the simulations are $m=3$, $M=10$, and $N=8000$. Each data point is
obtained as an average of $50$ independent runs.}
\end{figure}

Let $d(i,j)$ denotes the distance between node $i$ and node $j$, and
thus the average distance of the model with system size $N$ is
\begin{eqnarray}L(N)=\frac{2\sigma(N)}{N(N-1)}.\end{eqnarray}
where the total distance is
\begin{eqnarray}\sigma(N)=\sum_{1\leq i<j\leq N } d(i,j).\end{eqnarray}
Intuitively, when a new node is added, the distance between old
nodes will not increase. Hence we have
\begin{eqnarray}\sigma(N+1)\leq\sigma(N)+\sum_{i=1}^N d(i,N+1),\end{eqnarray}
thus
\begin{eqnarray}\sigma(N+1)\leq\sigma(N)+\sum_{i=1}^N d(i,x)+N,\end{eqnarray}
where $x$ is the active node connected to the newly added one. Since
$p$ is nonzero, by using mean-field approximation[26,27], we have
\begin{eqnarray}\sum_{i=1}^N d(i,x)\approx L(N)(N-1).\end{eqnarray}
Thus, the inequality (20) reduces to
\begin{eqnarray}\sigma(N+1)\leq\sigma(N)+\frac{2\sigma(N)}{N}+N.\end{eqnarray}
Rewriting (22) in continuous form will yield
\begin{eqnarray}\frac{d\sigma(N)}{dN}\leq\frac{2\sigma(N)}{N}+N,\end{eqnarray}
which leads to
\begin{eqnarray}\sigma(N)\leq N^2\ln N+B,\end{eqnarray}
where $B$ is a constant. As $\sigma(N)\sim N^2L(N)$ and $N$ is
sufficiently large, we obtain $L(N)\leq\ln N$, i. e., the increasing
tendency of $L(N)$ is not faster than $\ln N$, which predicts the
presence of small-world property.

In fact, the GKE network is similar to a chain of dense clusters
locally connected, i. e., it is like a regular lattice in
topological view. For this peculiar topology, all of the links in
the network are local. When a perturbation is introduced, the
network undergoes a cross-over from structured network to
unstructured network. Actually, the perturbation just means that,
with a probability, every node rewires one of its local links to a
randomly selected node, which is precisely the definition of the
model proposed by Watts and Strogatz[12]. That is to say, the
cross-over is just the small-world phase transition[24].

In Fig. 6, we show the dependence of average distance $L$ on system
size $N$ with $p=0.00$ and $p=0.01$ in GKE network. For $p=0.00$,
the average distance grows linearly $L\propto N$, the same behavior
observed in one-dimensional regular lattices. Once $p$ is a small
finite value, $L$ becomes logarithmic related to $N$, i. e.,
$L\propto \ln N$. The logarithmic increase of average distance with
system size predicts that the phase transition from large- to
small-world occurs, which is in agreement with the analytical
result.

\begin{figure}
\scalebox{0.9}[0.9]{\includegraphics{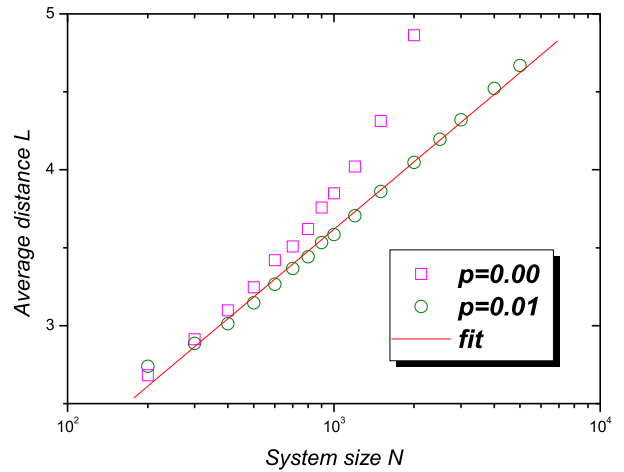}} \caption{(Color online)
Illustration of the average distance $L$ as a function of $N$, with
$p=0.00$ (squares) and $p=0.01$ (circles). When a perturbation
$p=0.01$ is introduced, $L$ grows logarithmically with $N$. The
values can be fitted well by a straight line, which is typical of
the small world effect. Other parameters for these simulations are
$m=4$ and $M=10$. Each data point is obtained as an average of $100$
independent runs. }
\end{figure}

\

It should be noted that, although we introduce a structural
perturbation into the network, the scale-free property is not
affected and the power-law exponent $\gamma=3$ is maintained.
Numerical simulations shown in Fig. 7 confirm this feature.

\begin{figure}
\scalebox{0.9}[0.9]{\includegraphics{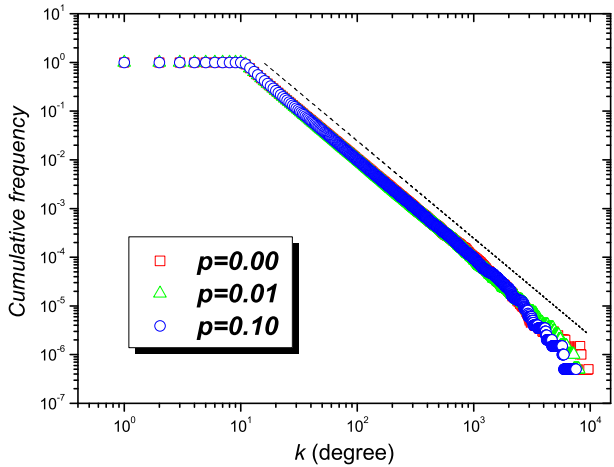}} \caption{(Color online)
Illustration of the cumulative degree distribution of the GKE
network with perturbation $p=0.00$ (squares), $0.01$ (upward
triangles), and $0.10$ (circles). The fitted power-law exponent is
$\gamma-1=1.97\pm0.02$. Other parameters for these simulations are
$m=10$, $M=20$, and $N=2\times10^4$. Each data point is obtained as
an average of $100$ independent runs. The dashed line has slope
$-2.0$ for comparison.}
\end{figure}

\subsection{Universal scaling behavior of clustering coefficient}
In the following subsection, we investigate the dependence of
clustering coefficient $C$ on perturbation parameter $p$. Analogous
to the derivation of clustering coefficient in GKE network without
perturbation, we give an approximately analytical result. According
to the modification of the model, when a new node $l$ with $m$ links
is added into the network, one of the links is attached to a
randomly selected inactive node $s$ with probability $p$. That is to
say, with probability $p$, one of the neighbors of $l$ is inactive.
Since the system size $N$ is large compared with $M$, we assume that
node $s$ is apart from the active nodes[28]. Thus, $m-1$ possible
links between neighbors of $l$ are missed. Furthermore, node $s$ is
always apart from the afterward added nodes that are connected to
node $l$, which causes another $k-m$ possible links missed. Thus we
have
\begin{eqnarray}C(l)&=&\frac{m-1\times p}{M}[1-\frac{(k_l-m+1)(k_l-m)}{k_l(k_l-1)}]\nonumber\\&-&p\frac{m-1\times p}{M}\frac{2(m-1)}{k_l(k_l-1)}\nonumber\\&-&p\frac{m-1\times p}{M}\frac{2(k_l-m)}{k_l(k_l-1)}.\end{eqnarray}
Similar to the derivation of Eq. (16), we have
\begin{eqnarray}C=\frac{m}{M}(\frac{5}{6}-\frac{7}{30m})-\frac{1}{M}(\frac{13}{6}-\frac{7}{30m})p+\mathcal{O}(p^2)\end{eqnarray}

It is worthwhile to note that the scaling behavior $C\sim1/M$ is
conserved though there exist certain fluctuations in the network
which lead to a structure transition. That is to say, $C\sim1/M$ is
a universal scaling behavior of clustering coefficient induced by
deactivation mechanism. Fig. 8 shows the log-log plot of the
clustering coefficient $C$ versus $M$ with different perturbation
parameters obtained by simulations. We can see that the perturbation
has almost no effect on the scaling behavior of $C$, which agrees
well with the analytical result.

From Eq. (26), we know that, when $p$ is sufficiently small, the
clustering coefficient $C$ has a linearly relation with $p$. Fig. 9
shows the simulation result of clustering coefficient $C$ as a
function of perturbation parameter $p$, with $m=4$ and $M=10$. The
slope found numerically is 0.223, slightly larger than the
analytical result $\frac{1}{M}(\frac{13}{6}-\frac{7}{30m})=0.21$.
The deviation is due to the approximation[28] used in the
theoretical derivation of $C$. It is clear that the node $s$ is not
always apart from all the active nodes, which actually causes less
than $m-1$ possible links missed between neighbors of $l$. Thus, we
can easily find the precise slope should be a little larger than
that obtained from Eq. (26).

\begin{figure}
\scalebox{0.9}[0.9]{\includegraphics{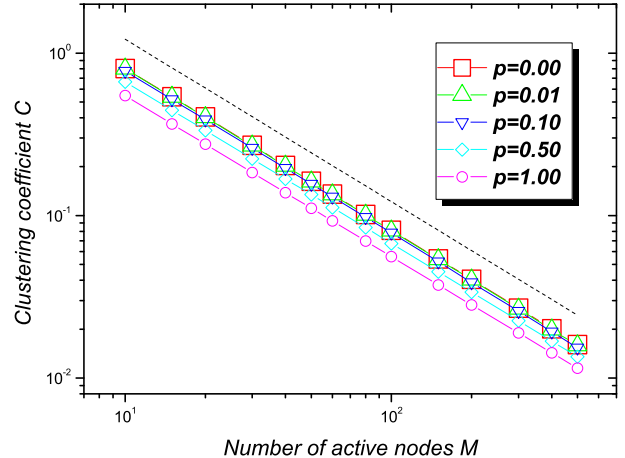}} \caption{(Color online)
Illustration of the clustering coefficient $C$ as a function of
number of active nodes $M$, with perturbation parameter $p=0.00$
(squares), $p=0.01$ (upward triangles), $p=0.10$ (downward
triangles), $p=0.50$ (diamonds), and $p=1.00$ (circles). The average
fit slope for the simulations is $0.994$. Other parameters for these
simulations are $m=10$ and $N=2\times10^4$. Each data point is
obtained as an average of $100$ independent runs. The dashed line
has slope $-1.0$ for comparison.}
\end{figure}

\begin{figure}
\scalebox{0.9}[0.9]{\includegraphics{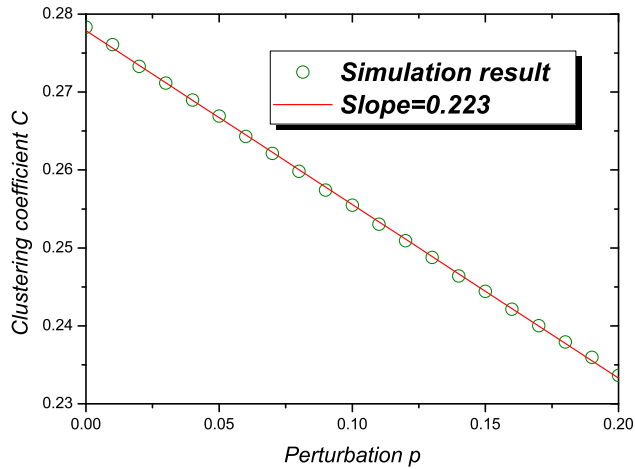}} \caption{(Color online)
Illustration of the clustering coefficient $C$ as a function of
perturbation parameter $p$. The fit slope is $0.223$. Other
parameters for this simulation are $m=4$, $M=10$, and
$N=2\times10^4$. Each data point is obtained as an average of $100$
independent runs.}
\end{figure}

\section{DISCUSSIONS AND CONCLUSIONS}
We first discuss the relationship between the numerical behavior of
clustering coefficient in the coauthor and paper citation networks
and our theoretical result. According to the model of Ref. [19],
each of the authors and papers is assigned a topic, and authors can
only cite, produce papers or coauthor with others in their own topic
area. It means that, by topics the whole network is divided into
many subnetworks which evolve separately and simultaneously. Each of
the subnetworks can be reduced to a GKE network and the number of
these GKE subnetworks is just the number of the topics denoted as
$n$. In each subnetwork, the number of \emph{active} authors who are
doing research or the number of \emph{active} papers that are likely
to be cited just corresponds to the number of active nodes in GKE
network, which is denoted as $M$. Since the whole network is divided
into $n$ subnetworks, we intuitively know that the number of
\emph{active} authors or papers in each subnetwork is inversely
proportional to the number of topics, i. e., $M\sim 1/n$. For each
subnetwork can be treated as a GKE network, incorporating with our
theoretical result $C\sim 1/M$, we can easily obtained that $C\sim
n$. Since each subnetwork evolves parallelly, the clustering
coefficient of the whole network has the same behavior that it is
linearly correlated with the number of topics. Therefore, by using
our theoretical result we can indicate that the numerical behavior
of clustering coefficient in the coauthor and paper citation network
presented in Ref. [19] is due to the deactivation mechanism.
Furthermore, in the above discussion, we reduce the aging mechanism
to deactivation mechanism. In fact, in the model of Ref. [19], the
aging effect is introduced by an aging function. To this point, we
conjecture that there might be similar scaling behaviors of $C$ in
networks generated by other forms of aging mechanism.

Finally, it is worthwhile to point out that, to our knowledge, no
empirical data are available to illustrate the theoretical scaling
behavior of clustering coefficient. Nevertheless, this interesting
property is due to the deactivation process which is a special case
of aging effect. In network evolution, aging is a universal
mechanism. Therefore, this simple theoretical result of $C$ will
have a rich practical significance and potential applications in
future network research. Meanwhile, such scaling behavior of $C$
should be given further considerations from empirical
investigations.

\

In summary, motivated by the aging effect governing the evolution of
the coauthor and paper citation networks, a generalized deactivation
model of network called GKE is presented in this paper. We study
analytically and by simulations several topological features of this
model, such as the degree distribution and clustering coefficient.
Most importantly, an interesting scaling behavior of the clustering
coefficient $C\sim 1/M$ is obtained, which shows that the numerical
result recently observed in the coauthor and paper citation networks
is due to deactivation mechanism. By introducing a perturbation, the
GKE network undergoes a small-world phase transition, while the
scaling behavior of $C$ is conserved. It indicates that $C\sim 1/M$
is a universal scaling behavior of clustering coefficient induced by
deactivation mechanism. In addition, we would like to emphasize that
our study unifies the concept of regular lattice, small-world graphs
and scale-free networks in a single model, and the GKE model
generalizes the new class of the networks with a crucial parameter
$M$.

Since the GKE networks present peculiar structure property, it will
be interesting to investigate the effect of their complex topology
features on the network dynamics[25,29-31]. Especially, the
clustering coefficient of GKE network is precisely tunable by
parameter $M$ or $p$ without changing the degree distribution.
Therefore, the model can be used to quantitatively study the effect
of clustering on network synchronization[32-34] and network
epidemics[31,35]. Research along this line is in progress.

\section*{ACKNOWLEDGMENTS}
Shi thanks the program for New Century Excellent Talents in
University of China for financial support (NECT-04-0510). This work
is also supported by National Natural Science Foundation of China
under Grant Nos. 70471084 and 10372045.


\begin{thebibliography} {1}
\bibitem{1}
R. Albert and A.-L. Barab\'asi, Rev. Mod. Phys. \textbf{74}, 47
(2002).
\bibitem{2}
S. N. Dorogrovtsev and J. F. F. Mendels, Adv. Phys. \textbf{51},
1079 (2002).
\bibitem{3}
M. E. J. Newman, SIAM Rev. \textbf{45}, 167 (2003).
\bibitem{4}
M. Faloutsos, P. Faloutsos and C. Faloutsos, Computer Communications
Review \textbf{29}, 251 (1999).
\bibitem{5}
R. Pastor-Satorras, A. Vazquez, and A. Vespignani, Phys. Rev. Lett.
\textbf{87}, 258701 (2001)
\bibitem{6}
B. A. Huberman, \emph{The Laws of the web} (MIT Press, Cambridge,
2001).
\bibitem{7}
R. Albert, H. Jeong and A. -L. Barab\'asi, Nature \textbf{401}, 130
(1999).
\bibitem{8}
H. Jeong, S. P. Mason, A.-L. Barab\'asi, and Z. N. Oltvai, Nature
\textbf{411}, 41 (2001).
\bibitem{9}
E. Ravasz, A. L. Somera, D. A. Mongru, Z. N. Oltvai, and A.-L.
Barab\'asi, Science \textbf{297}, 1551 (2002).
\bibitem{10}
S. L. Pimm, \emph{Food Webs} (University of Chicago Press, Chicago,
2002).
\bibitem{11}
S. Wasserman and K. Faust, \emph{Social Network Analysis} (Cambridge
University Press, Cambridge, 1994); J. Scott, \emph{Socical Network
Analysis: A Handbook} (Sage Publications, London, 2000).
\bibitem{12}
D. J. Watts and S. H. Strogatz, Nature \textbf{393}, 440 (1998).
\bibitem{13}
A. -L. Barab\'asi and R. Albert, Science \textbf{286}, 509 (1999).
\bibitem{14}
L. A. N. Amaral, A. Scala, M. Barth\'el\'emy, and H. E. Stanley,
Proc. Natl. Acad. Sci. U.S.A. \textbf{97}, 11 149 (2000).
\bibitem{15}
S. N. Dorogovtsev and J. F. F. Mendes, Phys. Rev. E \textbf{62},
1842 (2000).
\bibitem{16}
A. F. J. Van Raan, Scientometrics \textbf{47}, 347 (2000);
\bibitem{17}
H. Zhu, X. Wang, and J.-Y. Zhu, Phys. Rev. E \textbf{68}, 056121
(2003); K. B. Hajra and P. Sen, Phys. Rev. E \textbf{70}, 056103
(2004); K. B. Hajra and P. Sen, Physica A \textbf{346}, 44 (2005);
P.-Q. Jiang, B.-H. Wang, T. Zhou, Y.-D. Jin, Z.-Q. Fu, P.-L. Zhou,
and X.-S. Luo, Chin. Phys. Lett. \textbf{22}, 1285 (2005).
\bibitem{18}
K. Klemm and V. M. Egu\'iluz, Phys. Rev. E \textbf{65}, 036123
(2002).
\bibitem{19}
K. B\"orner, J. T. Maru, and R. L. Goldstone, Proc. Natl. Acad. Sci.
U.S.A. \textbf{101}, 5266 (2004).
\bibitem{20}
A. V\'azquez, M. Bogu\~n\'a, Y. Moreno, R. Pastor-Satorras, and A.
Vespignani, Phys. Rev. E \textbf{67}, 046111 (2003).
\bibitem{21}
C.-P. Zhu, S.-J. Xiong, Y.-J. Tian, N. Li, and K.-S. Jiang, Phys.
Rev. Lett. \textbf{92}, 218702 (2004).
\bibitem{22}
It means that each node is connected symmetrically to its $2z$
nearest neighbors. For detailed definition, see B. Bollob\'{a}s,
\emph{Random Graphs} (Academic Press, New York, 1985). We set $z=1$
in the simulations in this paper for the sake that the influence of
initial graph on the evolving of the network is minimal and can be
ignored.
\bibitem{23}
K. Klemm and V. M. Egu\'iluz, Phys. Rev. E \textbf{65}, 057102
(2002).
\bibitem{24}
M. E. J. Newman and D. J. Watts, Phys. Rev. E \textbf{60},
7332(1999).
\bibitem{25}
V. M. Egu\'iluz, E. Hern\'{a}ndez-Garc\'{i}a, O. Piro, and K. Klemm,
Phys. Rev. E \textbf{68}, 055102(R) (2003).
\bibitem{26}
T. Zhou, G. Yan, and B.-H. Wang, Phys. Rev. E \textbf{71}, 046141
(2005).
\bibitem{27}
Z.-M. Gu, T. Zhou, B.-H. Wang, G. Yan, C.-P. Zhu, and Z.-Q. Fu,
arXiv: cond-mat/0505175.
\bibitem{28}
As we know, when $p=0$, the topology of the GKE network is similar
to a one-dimensional lattice. Thus, if $p$ is sufficient small, this
approximation is reasonable.
\bibitem{29}
K. Klemm, V. M. Egu\'iluz, R. Toral, and M. SanMiguel, Phys. Rev. E
\textbf{67}, 026120 (2003)
\bibitem{30}
K. Suchecki, V. M. Egu\'iluz, and M. S. Miguel, Phys. Rev. E
\textbf{72}, 036132 (2005)
\bibitem{31}
V. M. Egu\'iluz and K. Klemm, Phys. Rev. Lett. \textbf{89}, 108701
(2002)
\bibitem{32}
P. N. McGraw, M. Menzinger, Phys. Rev. E \textbf{72}, 015101(R)
(2005).
\bibitem{33}
M. Zhao, T. Zhou, B. -H. Wang, G. Yan, and H. -J. Yang, preprint
arXiv: cond-mat/0510332.
\bibitem{34}
X. Wu, B. -H. Wang, T. Zhou, W. -X. Wang, M. Zhao, and H. -J. Yang,
Chin. Phys. Lett. 23, 1046 (2006).
\bibitem{35}
M. Bogu\~n\'a, R. Pastor-Satorras, and A. Vespignani, Phys. Rev.
Lett. \textbf{90}, 028701 (2003).







\end{thebibliography}
\end{document}